\documentclass[runningheads,a4paper]{llncs}

\usepackage{amsmath}
\usepackage{amssymb}
\usepackage{amsfonts}
\usepackage{makecell}            
\usepackage{graphicx}
\usepackage[dvipsnames]{xcolor}

\usepackage{url}
\urldef{\mailsa}\path|{k.michalewicz22, |   
\urldef{\mailsb}\path|m.barahona,|
\urldef{\mailsc}\path|b.bravi}@imperial.ac.uk|    
\newcommand{\keywords}[1]{\par\addvspace\baselineskip
\noindent\keywordname\enspace\ignorespaces#1}
\newcommand{\angstrom}{\mbox{\normalfont\AA}}

\begin{document}

\mainmatter  

\title{Machine learning approaches for interpretable antibody property prediction using structural data}

\titlerunning{Interpretable ML for antibodies}

%
%
\author{Kevin Michalewicz \and Mauricio Barahona \and Barbara Bravi}
%
\authorrunning{}

\institute{Department of Mathematics, Imperial College London, London SW7 2AZ, United Kingdom
\\
\mailsa
\mailsb
\mailsc\\}

%
%

\toctitle{Lecture Notes in Computer Science}
\tocauthor{Authors' Instructions}
\maketitle

\begin{abstract}
Understanding the relationship between antibody sequence, structure and function is essential for the design of antibody-based therapeutics and research tools. Recently, machine learning (ML) models mostly based on the application of large language models to sequence information have been developed to predict antibody properties. Yet there are open directions to incorporate structural information, not only to enhance prediction but also to offer insights into the underlying molecular mechanisms. This chapter provides an overview of these approaches and describes two ML frameworks that integrate structural data (via graph representations) with neural networks to predict properties of antibodies: ANTIPASTI predicts binding affinity (a global property) whereas INFUSSE predicts residue flexibility (a local property). We survey the principles underpinning these models; the ways in which they encode structural knowledge; and the strategies that can be used to extract biologically relevant statistical signals that can help discover and disentangle molecular determinants of the properties of interest.
\end{abstract}

\section{Introduction}
Antibodies are key actors in the adaptive immune system. Antibodies are proteins secreted by B cells that can bind with high affinity and specificity to cognate targets (\textit{antigens}) at specific sites 
(the \textit{epitope}). As such, antibodies are central to triggering the immune response and clearing infection, and are sought-after design targets for research and therapeutic applications. 

\begin{figure}
\centering
\includegraphics[clip, trim=0cm 26.55cm 19cm 0cm, width=\linewidth]{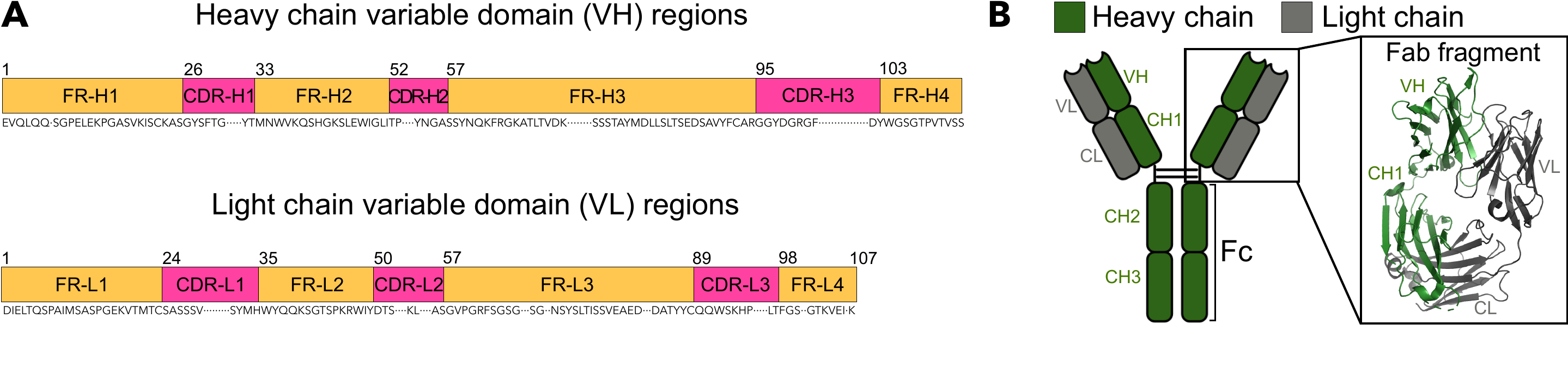}
\caption{\textbf{Antibody sequence and structure.} (\textbf{A}) Each variable domain of the heavy (VH) and light (VL) chains contains 3 CDRs and 4 FRs, defined via the Chothia position numbering scheme~\cite{Chothia1987}. (\textbf{B}) The structure of an antibody is Y-shaped, with two identical antigen binding fragments (Fab) and one \textit{crystallizable} fragment (Fc). Each Fab consists of the variable (VH and VL) and first constant domains (CH1 and CL), while the Fc comprises the CH2 and CH3 domains.}
\label{fig:1}
\end{figure}

Antibodies are Y-shaped proteins, where each tip of the Y is formed by a heavy and a light chain, each split into constant and variable regions (Figure~\ref{fig:1}). The variable domains carry the molecular determinants of antigen recognition, and each of its chains is organized into three highly variable complementarity-determining regions (CDRs) and four surrounding framework regions (FRs) that stabilize the structural scaffold~\cite{Janeway2001,Li2019}. CDRs constitute the majority of the antigen-binding site (the \textit{paratope}); among them, the CDR3 of the heavy chain (CDR-H3), often the main contributor to antigen binding~\cite{Chiu2019}, stands out for its sequence diversity, which is actively generated within single individuals by the VDJ recombination process~\cite{Mora2016}. 

Recently, there has been a surge in the development of computational antibody design tools aimed at selecting and optimizing candidate antibodies \textit{in silico} in terms of key properties such as binding affinity, binding specificity, solubility and stability~\cite{Seeliger2015,Jain2017,Sormanni2018}. These methods are critical to filter large candidate pools to prioritize variants most likely to succeed in downstream experimental validation. Machine Learning (ML) methods hold great promise in this setting, thanks to their success in protein-related prediction tasks~\cite{Alley2019,Yang2019,Senior2020}.

Among the properties that are a focus of computational modeling, binding affinity is essential to antibody design, as it governs the strength of antigen engagement, antigen/antibody dose requirements, as well as antibody developability trade-offs in multiple trait optimization protocols~\cite{Tabasinezhad2019,Wang2021,Makowski2022}.

Early antibody property prediction methods were primarily sequence-based, benefiting from large-scale datasets available at the sequence level~\cite{Sirin2016,Kang2021}. More recently, several ML models have incorporated both amino acid sequence and structural information for antibody property prediction~\cite{Sormanni2015,Wang2024}, including binding affinity and specificity~\cite{Kurumida2020,Yang2023}. Yet, despite the steadily expanding amount of structural data, including AlphaFold models~\cite{Jumper2021,Abramson2024}, the potential of structure-informed prediction remains underexploited. This is partly because training models on fully spatially resolved structures can be computationally intensive and data-hungry. Network models and geometric graphs, built from relative positions, distances and interactions within proteins, provide natural abstractions to account for structural information in a parsimonious yet effective way for protein modeling within ML pipelines~\cite{Delmotte2011,Qian2021}.

The importance of interpretability in ML for scientific inference is broadly appreciated~\cite{Rudin2021}. In the context of antibody discovery, models that are able to highlight the specific regions or interactions explaining their predictions offer valuable mechanistic insight, which in turn can strengthen model trustworthiness. The present chapter builds on this view, discussing ML principles and practices that encode protein structural knowledge via network or geometric graph constructions, while enabling biophysical interpretation in the context of antibody modeling. 

Protein property prediction tasks can be divided into \textit{global} or \textit{local}. Global properties describe a characteristic of the entire molecule, such as its stability or solubility, while local properties are defined at the level of individual structural units, such as individual atoms, residues or groups of residues. For example, single-residue properties include features like solvent accessibility~\cite{Durham2009}, conservation score~\cite{Landau2005}, secondary structure or local flexibility. Interestingly, local flexibility, which can be quantified through crystallographic B-factors, has been suggested to be linked to global properties such as binding affinity~\cite{Ovchinnikov2018}, as it shapes conformational selection and interface complementarity in antibody-antigen complexes~\cite{Teilum2009,Schneider2014,Carugo2018}.

We exemplify these two prediction categories through two ML structure-based methods that we have recently developed: ANTIPASTI~\cite{Michalewicz2024} and INFUSSE~\cite{Michalewicz2025}.
ANTIPASTI (ANTIbody Predictor of Affinity from STructural Information) exemplifies this philosophy for the prediction of a global property, binding affinity. It does so by creating representations of antibody-antigen complexes based on a dynamical interpretation of Elastic Network Models (ENMs) and processing them through a compact Convolutional Neural Network (CNN) that is made interpretable through model-agnostic and model-dependent strategies.
INFUSSE (Integrated Network Framework Unifying Structure and Sequence Embeddings) provides a complementary case study at the single-residue level by integrating sequence-level embeddings with a diffusive Graph Convolutional Network (diff-GCN) defined on geometric graphs to predict B-factors in antibody-antigen complexes.
%


The chapter is organized as follows. Section~\ref{sec:representaitons} introduces protein structure and sequence representations that serve as model inputs, including network abstractions, Normal Mode Analysis, and data-driven embeddings. Section~\ref{sec:prediction} briefly presents ANTIPASTI and INFUSSE model architectures and details how they leverage structural and sequence information for antibody property prediction. Section~\ref{sec:interpretability} discusses biophysical interpretability strategies, covering both model-agnostic and model-dependent approaches. Finally, Section~\ref{sec:conclusion} concludes with a summary and discussion.

\section{Protein Structure and Sequence Representations}\label{sec:representaitons}

\subsection{Network Models of Protein Structures}

The vastness of protein conformational space and associated computational cost make it challenging to model protein structures with full 3D spatial resolution at the atomistic level~\cite{Delmotte2011,Amor2016,Kmiecik2016}. Instead of an all-atom representation, it is possible to work with coarse-grained protein structure representations, focusing only on the heavy atoms or, most commonly, just one atom per amino acid, \textit{i.e.}, the $\alpha$-carbon atoms. Such coarse-grained, residue-level protein models have proven effective in various applications, including protein folding~\cite{Liwo2005,Kmiecik2012,Maisuradze2010}, aggregation studies~\cite{Nasica-Labouze2015}, protein-protein docking~\cite{Fleishman2010}, and protein flexibility prediction~\cite{Frezza2015}.

\paragraph{\normalfont{\textbf{Normal Mode Analysis.}}}
Well-established coarse-grained models of protein structures are formulated in terms of Elastic Network Models (ENMs)~\cite{Tirion1996,Bahar1997}, \textit{i.e.}, harmonic networks of residues. A prime example is the Anisotropic Network Model (ANM)~\cite{Atilgan2001}, which employs a simplified central potential but preserves the representation of anisotropic motions of the particles. Normal Mode Analysis (NMA) is then applied to ENMs to capture dynamical fluctuations around equilibrium
through a set of generalized coordinates (Normal Modes, NM)~\cite{Dykeman2010}. In our biomolecular modeling, NMs are useful to describe small-scale molecular deformations in relation to biological function~\cite{Levitt1985,Brooks1985,HenzlerWildman2007}. Although early applications of NMA were performed at the atomistic level, where each atom was treated as a particle~\cite{Go1983}, residue-level NMA on coarse-grained ENMs offers a computationally efficient alternative that captures collective motions relevant to protein function~\cite{Atilgan2001,Haliloglu1997,Hinsen1998}.

\begin{figure}
\centering
\includegraphics[clip, trim=0cm 3.25cm 1.6cm 0cm, width=\linewidth]
{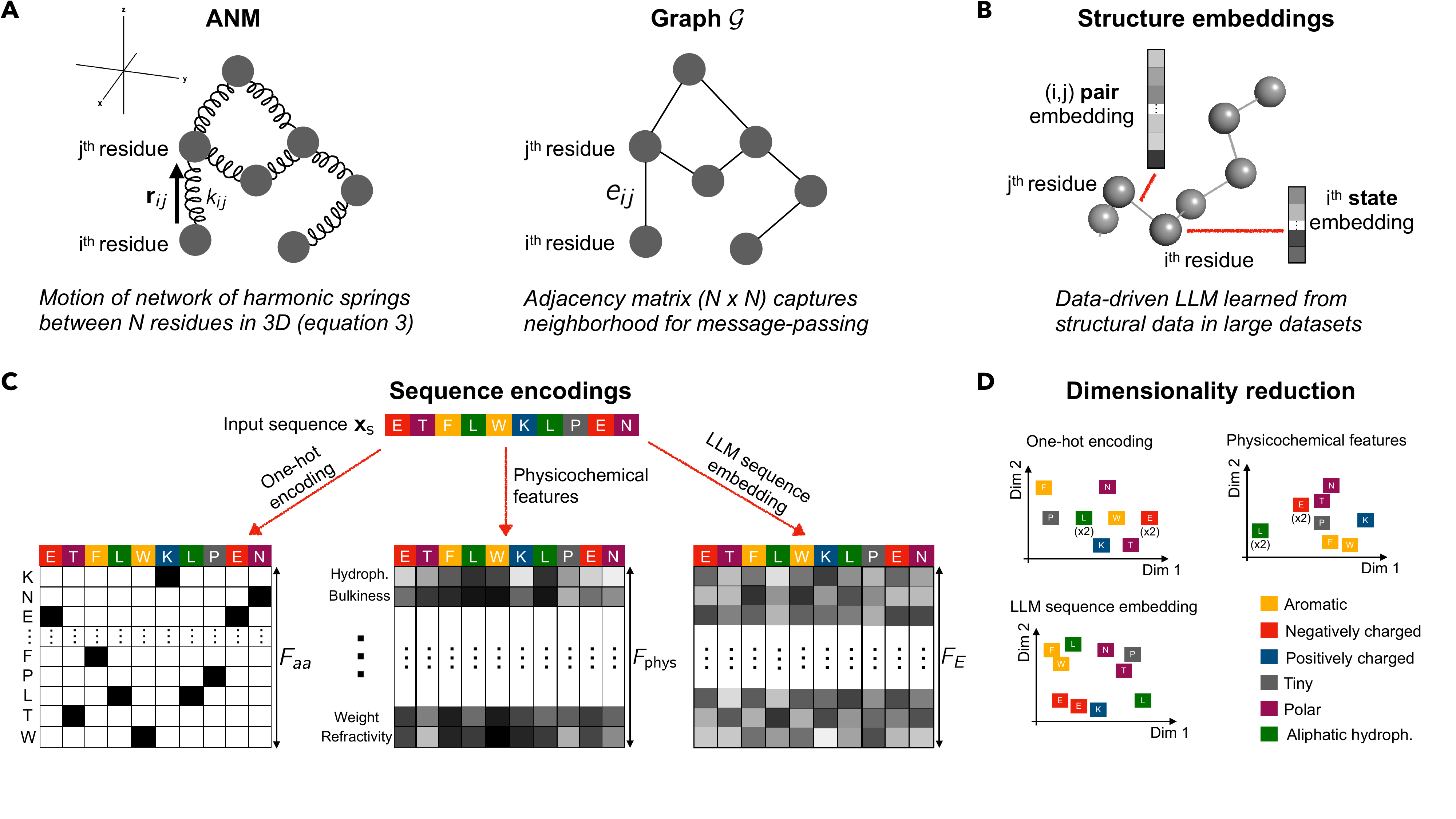}
\caption{\textbf{Protein representations.} (\textbf{A}) \textit{Left:} In ENMs, a protein structure is modeled as a network in which the nodes (typically residues) are connected by springs that represent the elastic force acting between them. For two connected nodes $i$ and $j$, the spring constant $k_{ij}$ quantifies the elastic interaction strength. The ANM is a particular case where $k_{ij} = k(r_{ij}^*)$, with $r_{ij}^*$
denoting the equilibrium distance between nodes $i$ and $j$. 
\textit{Right:} Graph representation of a protein structure as a network of $N$ nodes (typically residues) connected by edges $e_{ij}$ with weight $w(e_{ij})$ with adjacency matrix $A_{ij}=w(e_{ij})$. The diffusive dynamics and message-passing behavior of this system is governed by the associated graph Laplacian $\mathbf{L}$.
(\textbf{B}) Structure embeddings (data-driven and derived by LLMs) capture both per-residue (\textit{state}) and residue-residue (\textit{pair}) properties of the protein structure. (\textbf{C}) 
A protein sequence $\mathbf{x}_\mathrm{s}$ can be represented using one-hot encoding, physicochemical feature sets, or context-rich embeddings learned by an LLM. (\textbf{D}) Dimensionality reduction applied to the representations in C facilitate inspection through visualization in a lower-dimensional space. One-hot encodings do not capture similarity in physicochemical properties (denoted by colors) between residues. The $F_\text{phys}$ features quantifying physicochemical properties exhibit consistent groupings. The LLM-learned embeddings capture both biochemical similarity and context-dependence, so that the representation of a given residue is influenced by its physicochemical properties and by the sequence to which it belongs (context).}
\label{fig:2}
\end{figure}

Let $\mathbf{r} = [\mathbf{r}_1 \ldots \mathbf{r}_N]^\top \in \mathbb{R}^{3N}$ be the vector containing the 3D coordinates of the $\alpha$-carbons of the $N$ residues of a protein, and let $\mathbf{r}^*$ denote that vector at equilibrium (Figure~\ref{fig:2}A). The coarse-grained protein structure can then be viewed as a system of $N$ interacting particles under a potential energy dependent on the residue positions, $U(\mathbf{r})$.
Under small spatial displacements from equilibrium, the potential energy can be written (to second order) as:
\begin{equation} 
\label{eq:U_expansion}
U(\mathbf{r}) \simeq U(\mathbf{r}^{*}) + (\mathbf{r} - \mathbf{r}^{*})^\top \left. \nabla U(\mathbf{r})\right | _{\mathbf{r}^{*}} + \frac{1}{2} (\mathbf{r} - \mathbf{r}^{*})^\top 
\left. \nabla^2 U(\mathbf{r}) \right 
|_{\mathbf{r}^{*}} 
(\mathbf{r} - \mathbf{r}^{*})\;.
\end{equation}
 The potential energy of the equilibrium configuration, $U(\mathbf{r}^{*})$, can be shifted to zero without loss of generality~\cite{nmaBauer}, and the gradient term $ \left.\nabla U(\mathbf{r})\right|_{\mathbf{r}^{*}}$ also cancels since $\mathbf{r}^*$ is a local minimum of the potential energy. Therefore, only the second order term survives, and the potential energy close to equilibrium can be approximated as a harmonic potential, \textit{i.e.}, a quadratic form of the Hessian matrix $\nabla^2 U(\mathbf{r})\in\mathbb{R}^{3N\times 3N}$ evaluated at $\mathbf{r}^*$. 

The solutions to the equations of motion for small displacements of the residues under the quadratic potential (equation~\ref{eq:U_expansion}) take the form:
\begin{equation}
\boldsymbol{\delta}(t) := 
\mathbf{r}(t)-\mathbf{r}^{*}\propto\sum_{l=0}^{3N-1}C^{(l)}
\cos \left (\omega^{(l)}t+\phi^{(l)} \right) \, \mathbf{a}^{(l)}\;.
\end{equation}
In other words, the small displacements (or fluctuations) of the residues $\boldsymbol{\delta}(t)$ can be expressed as linear combinations of NMs indexed by $l$. Here, $C^{(l)}$ and $\phi^{(l)}$ are, respectively, the amplitude and phase of the $l^\text{th}$ NM, $\mathbf{a}^{(l)}\in \mathbb{R}^{3N}$ is the $l^\text{th}$ eigenvector of the Hessian matrix $\left. \nabla^2 U(\mathbf{r}) \right 
|_{\mathbf{r}^{*}}$ and $\omega^{(l)}$ is its associated eigenvalue, setting the natural frequency of the NM~\cite{Tama2001}. It is important to note that the first six NMs are trivial, as they are associated with rotational and translational invariances~\cite{Dubanevics2022}.

Ref.~\cite{Hinsen2005} proposed a particular form of the potential $U(\mathbf{r})$: 
\begin{equation}
\label{eq:amber-potential}
U(\mathbf{r}) = \sum_{\substack{i,j=1 \\ r_{ij}^{*}<\epsilon}}^N k(r_{ij}^{*})(r_{ij}-r_{ij}^{*})^2 \, ,\end{equation}
where $r_{ij}:=
\lVert\mathbf{r}_{ij} \rVert_2
=\lVert\mathbf{r}_i - \mathbf{r}_j \rVert_2$ is the Euclidean distance between 
particles $i$ and $j$, with $r_{ij}^*$ the distance at equilibrium; $\epsilon=7\angstrom$ is a cut-off radius; and $k(r_{ij}^{*})$ is the strength of the pairwise elastic force given by an $\alpha$-carbon force field derived by fitting to the Amber94 potential~\cite{Cornell1995}:
\begin{equation}
  k(r) =
    \begin{cases}
      8.6\times10^2 \, r-2.39\times10^3 & \text{ for } r<4\angstrom \\
      128\times10^4 \, r^{-6} & \text{ otherwise}
    \end{cases}       
\end{equation}
Using the NMs, the \textit{correlation of the fluctuations} of residues $i$ and $j$ can be computed as:
\begin{equation}
\label{eq:dccm}
X_{ij}:= \frac{\langle\boldsymbol{\delta}_i\cdot\boldsymbol{\delta}_j\rangle}{\langle\boldsymbol{\delta}_i^2\rangle^{1/2} \, \langle\boldsymbol{\delta}_j^2\rangle^{1/2}} \in [-1,1], 
\quad\text{with}\quad
\langle\boldsymbol{\delta}_i\cdot\boldsymbol{\delta}_j\rangle=k_BT\sum_{l=0}^{3N-1} \frac{{\mathbf{a}^{(l)}_i}\cdot\mathbf{a}^{(l)}_j}{{\omega^{(l)}}^2},\end{equation}
where $\cdot$ is the dot product in $\mathbb{R}^3$; the angular brackets 
indicate an ensemble average~\cite{Ichiye1991}; $k_B$ is the Boltzmann constant; $T$ is the absolute temperature; and $\mathbf{a}^{(l)}_i$ contains the three 
coordinates of the $l^\text{th}$ eigenvector of the Hessian matrix $\left. \nabla^2 U(\mathbf{r}) \right 
|_{\mathbf{r}^{*}}$ corresponding to residue $i$ (and similarly for $\mathbf{a}^{(l)}_j$). 
%
In this way, the correlations $X_{ij}$ provide a global representation of the effective linkage between residues mediated by the small-displacement dynamics of the protein.

\paragraph{\normalfont{\textbf{Graph-based representations}.}}
Protein structures at the residue level can also be represented more abstractly through an undirected graph $\mathcal{G}$ (Figure~\ref{fig:2}A), where each of the $N$ nodes represents an amino acid and $e_{ij}$, the edge connecting the nodes $i$ and $j$, has weight $w(e_{ij})$.
For unweighted graphs, $w(e_{ij})=1$ if there is an edge $e_{ij}$ and $w(e_{ij})=0$ otherwise. 
The adjacency matrix $\mathbf{A}\in\mathbb{R}^{N\times N}$ of $\mathcal{G}$ is symmetric with entries $A_{ij} = w(e_{ij})$. Another matrix associated with $\mathcal{G}$ is the $N \times N$ Laplacian matrix $\mathbf{L}:=\mathbf{D}-\mathbf{A}$, where $\mathbf{D} = \text{diag}(\mathbf{A} \, \mathbf{1})\in\mathbb{R}^{N\times N}$ and $\mathbf{1}$ is the $N \times 1$ vector of ones (\textit{i.e.}, $\mathbf{D}$ contains the node degrees on the diagonal). For undirected graphs, $\mathbf{L}$ is symmetric and positive semi-definite~\cite{Stehlík2017}. The graph Laplacian matrix is a central concept in spectral graph theory and is directly linked to diffusive processes on graphs, as its eigenvalues determine how fast such processes converge to equilibrium~\cite{Chung1997,Lambiotte2014}. It has also been shown that, although simplified, such protein graph representations capture information about geometric neighborhoods and weighted local energetic interactions, which can deliver important functional and structural insights~\cite{Delmotte2011,Amor2016}.

The simplest graph construction for protein structures is the geometric 
$\epsilon$-ball graph~\cite{Liu2020}, whereby residues within an $\epsilon$ distance of each other are assigned an unweighted edge. This graph has the following adjacency matrix:
\begin{align}
   A_{ij} =
    \begin{cases}
      1 & \text{ if } r_{ij} < \epsilon\\
      0 & \text{ otherwise}
    \end{cases}      
    \label{eq:GNM}
\end{align}
The cutoff is typically chosen in the range $\epsilon\in[8\angstrom,10\angstrom]$, following Ref.~\cite{Baldi2003}.



An alternative geometric construction is a simple \textit{weighted} Gaussian graph with (full) weighted adjacency matrix given by:
\begin{align}
 A_{ij}=\exp\left(
 -r_{ij}^2/\eta^2
 \right),  
 \label{eq:Gaussiangraph}
\end{align}
with length scale $\eta=8\angstrom$ as in Ref.~\cite{Opron2016}.

More sophisticated approaches construct atomistic graphs with energy-weighted edges (reflecting different chemical bonds and physicochemical interactions), which can be subsequently coarse-grained to the residue level~\cite{Delmotte2011,Song2021}.

\paragraph{\normalfont{\textbf{Machine-learned structure embeddings}.}}
Recently, ML models trained on a very large portion of the Protein Data Bank (PDB,~\cite{Berman2000}) have been leveraged to encode structures into fixed-size embeddings, producing high-dimensional residue-level representations that are enriched with spatial information~\cite{Lau2024}. Such structure-aware embeddings have been applied to tasks including structure similarity assessment~\cite{Kandathil2025}, structure searching~\cite{Greener2024}, property prediction~\cite{Blaabjerg2024,Danner2025}, and domain classification~\cite{Lau2024}. 
Transformer-based models for protein modeling like Refs.~\cite{Jumper2021,Abramson2024,Baek2021,Watson2023} learn \textit{state} and \textit{pair} embeddings, which correspond to per-residue and residue-residue properties of the protein structure, respectively (see Figure~\ref{fig:2}B). Structure embeddings of this type can be integrated into graph-based frameworks, as they encode geometric and interaction context analogous to biophysical descriptors, yet learned directly from data~\cite{Jing2021}.

\subsection{Sequence Encodings}
Protein sequences can be represented in several ways to serve as input for ML models (see Figure~\ref{fig:2}C for a schematic summary). The most basic representation, which does not consider relationships between residues, is \textit{one-hot encoding}, a binary vector representation in which each amino acid is given a unique index in a fixed dictionary of size $F_{aa}$~\cite{ElAbd2020} (typically $F_{aa}=20$, corresponding to the standard amino acid types). 
Given a sequence $\mathbf{x}_{\mathrm{s}}$ of length $N$, $\mathbf{\Phi}(\mathbf{x}_{\mathrm{s}})\in\lbrace{0,1\rbrace}^{N\times F_{aa}}
$ is called the one-hot encoded version of $\mathbf{x}_{\mathrm{s}}$.

The biochemical information carried by sequences is also routinely described by considering a set of physicochemical properties for each amino acid, 
such as aromaticity~\cite{Ofer2015}, hydrophobicity~\cite{Harding-Larsen2024,Li2024} or composite measures like Kidera factors~\cite{Rackovsky2010}. Beyond commonly used scales, hundreds of additional descriptors can be generated, covering properties such as mutability, free energy and steric bulk. These have been classified into several categories: conformation, energy, composition, volume, structure-activity, shape and polarity~\cite{Breimann2024}. The drawback of hand-picked feature approaches is that the chosen properties may not be the most relevant or informative for the specific task under consideration~\cite{Yang2018}. Another option is to represent sequences at the \textit{atomic level} via SMILES~\cite{Weininger1988}, a line notation that encodes molecules as strings that specify atom types and bond connectivity.

Although the above representations (\textit{embeddings}) are interpretable, they are semantically limited: one-hot encodings ignore residue-residue relationships and lack biochemical context, while physicochemical feature sets depend heavily on the quality and appropriateness of the features chosen for the representation. To overcome these limitations, there has been a recent surge in the use of advanced deep learning methods based on Large Language Models (LLMs) to create data-driven \textit{sequence embeddings} that capture complex statistical dependencies between residues along the sequence obtained from training on very large datasets.
Indeed, LLMs specifically tailored to proteins and trained in a self-supervised way~\cite{Raina2007} can learn, through their intermediate layers, high-dimensional sequence embeddings that are shown to recapitulate evolutionary, structural and functional signals (see Ref.~\cite{Brandes2022} and Figure~\ref{fig:2}C). Examples include ProtBERT~\cite{Elnaggar2022}, ESM~\cite{Rives2021}, and ProGen~\cite{Madani2023}. 
For a protein sequence $\mathbf{x}_{\mathrm{s}}$ of length $N$, we denote the embedding given by a hidden layer of a pre-trained LLM as $\mathbf{E}_{\mathrm{s}}\in\mathbb{R}^{N\times F_E}$ with $F_E$ being the embedding dimension (\textit{e.g.}, $F_E=1024$ for ProtBERT~\cite{Elnaggar2022}). Unlike the interpretable but semantically limited representations discussed above, LLM-derived embeddings are not directly interpretable in physicochemical terms, yet they encode much richer information learned from large-scale sequence statistics, enabling meaningful comparisons across sequences.
Each residue in the sequence is represented by a context-dependent embedding, meaning that the same amino acid type acquires a different representation depending on its surrounding sequence environment. This allows such embeddings to capture functional, evolutionary, and sequence-context information beyond the simple amino acid identity, as it typically becomes clear when visualized in reduced dimensions (Figure~\ref{fig:2}D). These high-dimensional, semantically rich representations thus provide residue-level features that can be directly used when modeling proteins as graphs.


However, although such sequence models have shown their utility and power for many tasks, they tend to perform accurately only when there are well-defined and statistically strong signals of residue covariation that are informative about the structural and functional properties modeled. Antibody interfaces, on the other hand, involve mainly the CDRs, whose behavior is hard to infer from sequence alone due to the fact that they are regions with high amino acid composition variability and conformational flexibility acting as binding interfaces in contact with external structural motifs (epitope). All these properties together make it difficult to rely on sequence-only methods and motivate the need for including and encoding structural information into our descriptions, \textit{e.g.}, as graphs or network protein representations. 

In ANTIPASTI, inputs are \textit{dynamics-aware} residue-residue correlation maps derived from normal modes of an ANM (equation~\ref{eq:dccm}). In INFUSSE, the graph is \textit{geometry-driven}, defined either by an $\epsilon$-ball graph (equation~\ref{eq:GNM}) or a smoothly decaying weighted-Gaussian adjacency 
(equation~\ref{eq:Gaussiangraph}). Both networks of residues represent different types of structural signals: ANTIPASTI's input maps capture patterns of correlated fluctuations, which involve global dynamical properties of the network, whereas INFUSSE is based on spatial proximity 
(representing packing~\cite{Halle2002} or biophysical interactions~\cite{Song2021}) as the basis for propagating node features into the graph neighborhood via message-passing.

\section{Integrating Structural Information for Predictive Modeling of Antibody Properties}\label{sec:prediction}


We now introduce the architectures of ANTIPASTI and INFUSSE as two examples of ML pipelines that integrate structural information for the prediction of, respectively, global and local protein properties. While exemplified here with \textit{regression} tasks, \textit{i.e.}, prediction of continuous properties, the techniques can be adapted to classification tasks, \textit{i.e.}, prediction of discrete class labels. 

\subsection{ANTIPASTI for Binding Affinity Prediction: An Example of Global Property Prediction}

Consider the following general regression problem. Given the input-output pairs $(\mathbf{X}^{(i)}, y^{(i)})_{i=1}^{N_{\text{s}}} \subseteq \mathcal{X}\times\mathbb{R}$, 
the aim is to learn a function $f\in \mathcal{F}$ such that it captures the input-output relationship, \textit{i.e.}, $f(\mathbf{X}^{(i)}) \approx y^{(i)}$, where $N_s$ is the number of available data entries and $\mathcal{F}$ is the model class, \textit{e.g.}, neural networks. 
For 2D image data, for instance, $\mathcal{X}=\mathbb{R}^{N\times N}$, where $N^2$ is the number of pixels.

\begin{figure}
\centering
\includegraphics[clip, trim=0cm 3.25cm 0.5cm 0cm, width=\linewidth]{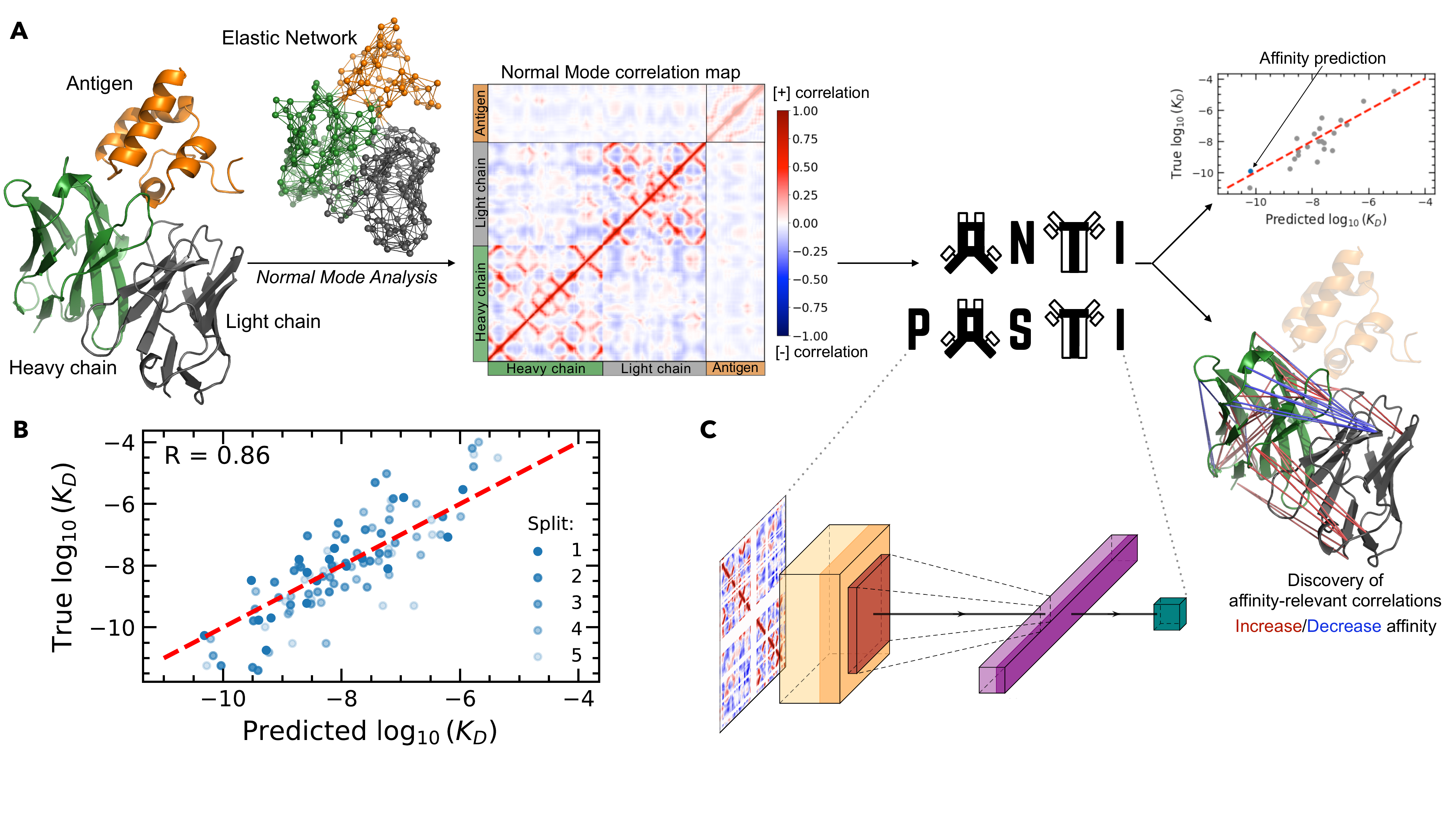}
\caption{\textbf{Overview of ANTIPASTI.} (\textbf{A}) From the PDB structure of the antibody-antigen complex, an elastic network model representation is created and a map of correlations between residues is calculated from normal modes, which carry the antigen's imprint on the antibody residues. Gaps are then added in place of absent residues through an alignment, and the antigen residues are removed from the correlation map to produce the input image to the ANTIPASTI's CNN architecture (see \textbf{C}). ANTIPASTI processes this input image and yields a binding affinity prediction and a map of affinity-relevant correlations. (\textbf{B}) ANTIPASTI predictions on the test set are compared to the ground-truth affinities for five training/test splits. (\textbf{C}) ML architecture of ANTIPASTI. Figure adapted from Ref.~\cite{Michalewicz2024}.}
\label{fig:3}
\end{figure}

ANTIPASTI trains a CNN in a supervised fashion (see Figure~\ref{fig:3}A,C) on input pairs given by: (i) NM correlation maps $\mathbf{X}\in\mathbb{R}^{N\times N}$ (equation~\ref{eq:dccm}), computed from antibody-antigen structures, and (ii) experimentally measured values of the binding dissociation constant ($K_D$). The model comprises a bank of $n_{\text{f}}$ $k\times k$ convolutional filters, followed by a ReLU activation function and a $p\times p$ max pooling layer with stride $s$. 
The latter is followed by a bias-free fully-connected layer from which the ANTIPASTI's global prediction, the value of $\log_{10}(K_D)$, is computed as follows. 

Let $\mathbf{z}\in\mathbb{R}^M$ be the flattened input of the ANTIPASTI fully-connected layer, produced by the network's convolutional layer (with $n_{\text{f}}$ filters and the non-linear activation) and the max pooling layer applied to a NM correlation map $\mathbf{X}$. Let $\mathbf{w}\in\mathbb{R}^M$ be the weights of the fully-connected layer. Then, since there is no bias, the predicted value of $K_D$ (denoted by $\hat{K}_D$) is obtained simply from:
\begin{equation}
\label{eq:ba}
\log_{10}(\hat{K}_D) = \mathbf{w}^\top\mathbf{z}=\sum_{i=0}^{M-1}w_i z_i, 
\quad \text{with} \quad M=n_{\text{f}} \,\left\lfloor\frac{N-k+1-p}{s}+1\right\rfloor^2 \, .
\end{equation}
The value of $M$ arises because every convolution operation yields a matrix of shape $(N-k+1)\times(N-k+1)$ which is 
downsized through max pooling with kernel shape $p\times p$ and stride $s$.

Convolutions are powerful when dealing with protein structures due to their robustness to detect features regardless of their exact position. This makes it possible to find common structural motifs relevant to protein function~\cite{Singh2003,Alzubaidi2021}. Further, given the multiscale nature of proteins, the hierarchical architecture of CNNs is ideal for capturing features at different levels of resolution~\cite{Delmotte2011,Soleymani2018}.

Antibody-antigen binding typically follows an \textit{induced fit} mechanism, characterized by small, localized arrangements such as CDR loop adjustments or slight domain reorientations~\cite{Kourentzi2008,SelaCulang2012,Barozet2018}. As NMA captures small-amplitude displacements around the equilibrium position, it is well-suited for binding affinity prediction, a relationship also proposed in Refs.~\cite{nmaBauer,Lee2006}. The induced fit model inherently assumes small structural adjustments and is thus embedded in the choice of the harmonic ANM potential energy (equation~\ref{eq:amber-potential}), which serves as a biophysical prior for the ML model.

After learning the CNN model, we tested the accuracy of ANTIPASTI for prediction of the binding affinity constant $K_D$ on curated, unseen test datasets from SAbDab~\cite{Dunbar2014,Schneider2022}. 
Figure~\ref{fig:3}B shows the values $\hat{K}_D$ predicted by the CNN with pooling on an unseen test set for five training/test splits compared to the true values $K_D$, achieving a Pearson correlation of $R=0.86$ when combining data across splits. The random splits were prepared robustly to ensure there is less than 90\% antibody sequence identity and that the antigens are different between the training and test data. 

For comparison, a linear regression approach to predicting $K_D$ from NMs achieves a mean $R=0.64$ (see Ref.~\cite{Michalewicz2024}), substantially lower than ANTIPASTI, indicating that it is crucial to include non-linearities and layer types with robust feature detection properties, such as convolutional layers.

\subsection{INFUSSE for B-factor Prediction: An Example of Single-Residue Property Prediction}

In contrast to the prediction of a global protein property achieved by ANTIPASTI, we now consider the prediction of a residue-level property. This can be formulated as a graph node regression task, where each node of the graph is associated with a residue. 

For a protein with $N$ residues, let $\mathbf{X}\in\mathbb{R}^{N\times F}$ be a matrix that contains $F$ features for each residue, \textit{e.g.}, a one-hot encoding $\mathbf{\Phi}(\mathbf{x}_{\mathrm{s}})$ or a residue-level embedding obtained from learned representations $\mathbf{E}_{\mathrm{s}}$ of the protein sequence $\mathbf{x}_{\mathrm{s}}$ (Figure~\ref{fig:2}C). 
Let $\mathbf{L}\in\mathbb{R}^{N\times N}$ be the Laplacian matrix of the graph $\mathcal{G}$, derived from the atomic coordinates of the residues, \textit{e.g.}, the Cartesian coordinates of the $N$ $\alpha$-carbons $\mathbf{r}\in\mathbb{R}^{3N}$, which can be found in PDB files. Finally, let $\mathbf{B}\in\mathbb{R}^N$ be a vector containing the target values that we aim to predict for each node, \textit{i.e.}, at the local residue level. 

The INFUSSE architecture consists of a \textit{sequence block} and a \textit{graph block} that are learned sequentially and combined to produce the predictions:
\begin{equation}\label{eq:infusse}\hat{\mathbf{B}}:=\text{INFUSSE}(\mathbf{x}_{\mathrm{s}}, \mathbf{r}):= S_{\mathrm{block}} \left(\mathbf{x}_{\mathrm{s}}, \mathbf{E}_{\mathrm{s}} \right) +
G_{\mathrm{block}}(\mathbf{r}, \mathbf{X} (\mathbf{x}_{\mathrm{s}}, \mathbf{E}_{\mathrm{s}})
)\;.
\end{equation}

The sequence block $S_{\mathrm{block}}$ integrates one-hot encoded residue information and LLM embeddings through a series of learnable transformations that produce an enriched residue-level representation fine-tuned for the prediction task.

The graph block corresponds to a two-layer diff-GCN~\cite{Peach2020}: 
$G_{\mathrm{block}}(\mathbf{r}, \mathbf{X}):= \text{diff-GCN}_t(\mathbf{L}(\mathbf{r}),\mathbf{X}
   ):=e^{-t \mathbf{ L}} \, \text{ReLU}\left(e^{-t \mathbf{ L}} \mathbf{X} \,  \mathbf{W}^{(0)}\right) \mathbf{W}^{(1)}$, where $\mathbf{W}^{(l)} \in \mathbb{R}^{F_{l+1} \times F_l}$ is the matrix of learnable weights for the $l^{\text{th}}$ layer of the diff-GCN. Note that the matrix $e^{-t \, \mathbf{L}}$ is the diffusion transition associated with the graph Laplacian $\mathbf{L}$, and $t\in\mathbb{R}_{>0}$ is a learnable scale parameter.

GCNs differ fundamentally from CNNs. Instead of applying spatially localized filters, GCNs update node features via a message-passing scheme derived from a first-order approximation of spectral graph convolutions~\cite{Kipf2017,Gilmer2017}, such that each node aggregates information from its neighbors. In the diff-GCN variant, this propagation is further modulated by the graph diffusion operator $e^{-t \mathbf{ L}}$, where the parameter $t$ controls the effective range of information flow.

\begin{figure}
\centering
\includegraphics[clip, trim=0cm 3.25cm 10cm 0cm, width=\linewidth]{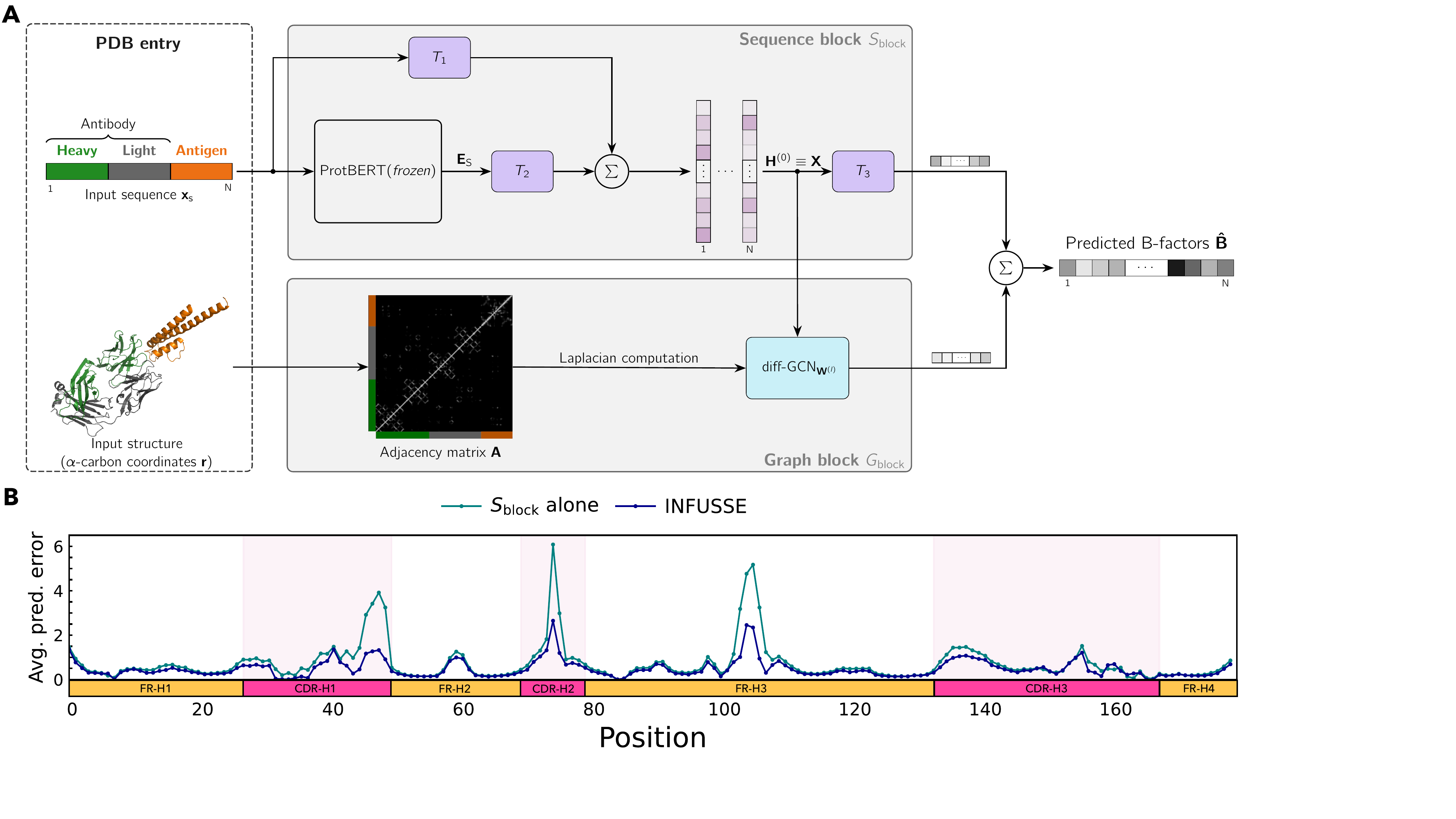}
\caption{\textbf{Overview of INFUSSE.} (\textbf{A}) In the INFUSSE architecture, a frozen LLM (ProtBERT) and a diffusive Graph Convolutional Network (diff-GCN) are combined to predict B-factors for antibody-antigen complexes. In the sequence block ($S_{\mathrm{block}}$), input sequences are encoded by the frozen ProtBERT model and passed through a learnable non-linear layer $T_2$, then summed with their one-hot encoded version transformed by another learnable non-linear layer $T_1$, producing enriched sequence embeddings $\mathbf{X}$ that are further transformed by a learnable non-linear layer $T_3$. The resulting representations are summed with the output of the graph block $G_{\mathrm{block}}$, \textit{i.e.}, of the diff-GCN with learnable parameters $t$, $\mathbf{W}^{(0)}$ and $\mathbf{W}^{(1)}$ that takes $\mathbf{X}$ from $S_{\mathrm{block}}$ as input node features and leverages the Laplacian of a geometric graph constructed from the antibody-antigen structure. (\textbf{B}) Prediction errors of INFUSSE, $\varepsilon_{\text{INFUSSE}, n}^{(q)}$, and $S_{\mathrm{block}}$ alone, $\varepsilon_{S_\mathrm{block}, n}^{(q)}$, for each position $n$ of the heavy chain antibody variable region averaged over the samples $q \in \mathcal{Q}$. Figure adapted from~\cite{Michalewicz2025}.}
\label{fig:4}
\end{figure}

We have applied the INFUSSE model (Figure~\ref{fig:4}A) to the prediction of 
experimental standardized B-factors, 
which can also be extracted directly from PDB entries~\cite{Michalewicz2025}.
To train INFUSSE, we retrieved antibody-antigen complexes and unbound antibodies with B-factor annotation. INFUSSE is trained in a supervised way by minimizing the mean squared error of B-factor prediction in a \textit{two-step} manner. In a first step, we train the sequence block $S_{\mathrm{block}}$ by learning the parameters of $T_1$, $T_2$ and $T_3$ to predict B-factors from sequence alone, \textit{i.e.}, from the sequence information $\mathbf{x}_{\mathrm{s}}$ and the LLM (ProtBERT) embedding $\mathbf{E}_{\mathrm{s}}$. In a second step, we incorporate the graph block $G_{\mathrm{block}}$ and we optimize the parameters of the diff-GCN ($t$, $\mathbf{W}^{(l)}$) \textit{jointly} with the ones of $T_1, T_2, T_3$ (initialized to the values obtained from the first step). Note that ProtBERT is kept frozen, and is only used to compute informative sequence embeddings $\mathbf{E}_{\mathrm{s}}$ used as input.

After training, the model was evaluated on the test set by computing the Pearson correlation coefficient $R$ between the ground truth $\mathbf{B}$ and predicted standardized B-factors $\hat{\mathbf{B}}$, repeating training and testing for 10 independent training/test data splits. INFUSSE achieved $R=0.71 \pm 0.01$, averaged over the 10 training/test splits, for the prediction of standardized B-factors (see Table~\ref{table:gcn-bf-perf}). This performance was achieved with a weighted Gaussian graph~(equation~\ref{eq:Gaussiangraph} with $\eta=8\angstrom$), but other similar setups performed similarly, confirming 
the robustness of the INFUSSE architecture.

Table~\ref{table:gcn-bf-perf} shows the results of two other models that contain only \textit{part} of the full INFUSSE architecture.
The `$S_{\mathrm{block}}$ alone' model is our `step 1' model trained only on sequence and ProtBERT embeddings, \textit{i.e.}, $S_{\mathrm{block}} (\mathbf{x}_{\mathrm{s}}, 
 \mathbf{E}_{\mathrm{s}} )$, and it achieved an average $R=0.64$. The `$S_{\mathrm{block}}$ (no LLM embeddings) + $G_{\mathrm{block}}$' model corresponds to $S_{\mathrm{block}} \left(\mathbf{x}_{\mathrm{s}}, \mathbf{0} \right) +
G_{\mathrm{block}}(\mathbf{r}, \mathbf{X} (\mathbf{x}_{\mathrm{s}}, \mathbf{0})
)$, \textit{i.e.}, one-hot encodings are used in place of ProtBERT embeddings, and it achieved an average $R=0.55$.
Altogether, these results underscore the importance of including both sequence embeddings and structural graphs within INFUSSE for the prediction of single-residue properties that are influenced by local residue neighborhoods. 

In Table~\ref{table:gcn-bf-perf}, we also compared INFUSSE with two models from the literature that perform well on B-factor prediction for generic proteins.
Although the state-of-the-art (SOTA) Long Short-Term Memory (LSTM) model has been shown to achieve $R$ between $0.6-0.8$ on generic protein benchmarks~\cite{Pandey2023,Bramer2018}, it only produced an average $R=0.48$ when applied to our antibody-antigen dataset. This reflects the difficulty of predicting B-factors for antibodies due to their high sequence variability and the presence of unstructured regions. 
Similarly, a baseline graph model that relies on the pseudoinverse of the graph Laplacian with no learning~\cite{Bahar2010} produced $R= 0.01$ for both weighted Gaussian and $\epsilon$-ball graphs on our antibody dataset, in contrast to $R$ between $0.65$ and $0.8$ on generic proteins~\cite{Rader2005}. Again, this observation underscores the challenges of B-factor prediction in antibody-antigen complexes, and the need to combine and leverage jointly enriched sequence representations and structural graph information, as in our INFUSSE model. 

\begin{table}
\centering
\footnotesize
\caption{\textbf{Performance in residue B-factor prediction for antibody-antigen complexes.} The measure of performance reported for different methods is the Pearson correlation coefficient $R$ between experimental and predicted B-factors of the test set, averaged over 10 training/test splits, with standard deviation across splits. Table adapted from~\cite{Michalewicz2025}.}
\resizebox{\textwidth}{!}{
\begin{tabular}{|l|c|c|c|}
\hline
\textbf{Method} & \textbf{\makecell{Sequence \\representation}} & \textbf{\makecell{Structure \\representation}} & $R$ \\ \hline
INFUSSE ($S_{\mathrm{block}}$ + $G_{\mathrm{block}}$) & \makecell{One-hot encoding \\ \& LLM embeddings} & Weighted Gaussian graph & $0.71\pm 0.01$ \\ \hline
$S_{\mathrm{block}}$ alone (no structure) & \makecell{One-hot encoding \\ \& LLM embeddings} & --- & $0.64\pm 0.02$ \\ \hline
$S_{\mathrm{block}}$ (no LLM embeddings) + $G_{\mathrm{block}}$ & One-hot encoding & Weighted Gaussian graph & $0.55\pm 0.04$ \\ \hline \hline
LSTM~\cite{Pandey2023} (SOTA for general proteins) & One-hot encoding & \makecell{Raw coordinates $\mathbf{r}$, secondary \\structure, and chain breaks} & $0.48\pm 0.06$ \\ \hline
Laplacian pseudoinverse (no learning) & --- & Weighted Gaussian graph & $0.01\pm 0.04$ \\ \hline
\end{tabular}}
\label{table:gcn-bf-perf}
\end{table}

\section{Biophysical Interpretability}\label{sec:interpretability}

Interpretability in ML seeks to explain model predictions, both in connection to its internal workings and to the modeled system, \textit{e.g.}, by relating variations in inputs and learned representations to variations in outputs in terms of intelligible, meaningful features or variables.
Ref.~\cite{Ras2018} spells out the desiderata of interpretability approaches, emphasizing fidelity to the original input-output mapping, clarity, and parsimony of the explanation. In practice, a simple model such as linear regression is intrinsically explainable insofar as the model's prediction depends in a clear and transparent way on a weighted sum of features, with each regression coefficient reflecting the importance of the corresponding feature. The prediction by a decision tree can be rationalized in terms of a set of decisions in the feature space, and a parsimonious summary of the predictive contribution of each feature can be obtained via metrics of feature importance~\cite{Breiman2001}.

On the other hand, deep neural networks typically trade algorithmic transparency for performance, requiring \textit{post hoc} interpretability methods to ascribe performance to particular variables or, alternatively, by constructing architectures that are interpretable by design~\cite{Abdullah2021}. 

\subsection{Model-Agnostic Interpretability Strategies: Feature Importance}
Interpretability approaches can be broadly classified as either \textit{model-agnostic} or \textit{model-dependent}. Within the first group, different approaches are typically employed to compute concise, \textit{post hoc} metrics of feature importance through \textit{importance factors}, which measure how strongly an input variable (feature) influences the predictive performance of a model~\cite{Schossler2024}.
LIME (Local Interpretable Model-agnostic Explanations) fits a simple linear regression model around a prediction and uses the regression coefficients to quantify feature importance~\cite{Ribeiro2016}, whereas SHAP (SHapley Additive exPlanations) 
assigns each feature a Shapley value that reflects its expected marginal contribution, thereby providing an information-theoretic decomposition of the model's output into contributions from each input variable~\cite{Lundberg2017}. 

These methods have been applied in the context of immune proteins. Examples include the prediction of epitope-specific T-cell receptors with random forest models, where Gini-based importance factors rank discriminatory $\beta$-chain CDR3 positional and physicochemical features~\cite{DeNeuter2018}, and a \textit{post hoc} pipeline based on the model-agnostic interpretability strategy of ``anchors''~\cite{Ribeiro2018} to extract human-readable, specificity-determining binding motifs in T-cell receptors~\cite{Papadopoulou2022}. Additional examples of ML interpretability strategies applied to molecular immunology are discussed in Ref.~\cite{Bravi2024}.

\paragraph{\normalfont{\textbf{Antibody region importance for binding affinity.}}} 

Model-agnostic interpretability strategies have been used within ANTIPASTI to introduce a region-level importance factor that quantifies the change in the prediction error when a given region is excluded from the final affinity estimation. The importance $I$ for a region $s^*$ is calculated as:
\begin{equation}
\label{eq:importance_factor}
I(s^*)=100\frac{\bar{N}(s_{\text{Best}})}{\bar{N}(s^*)}\frac{\bigg\rvert\text{MSE}\big\rvert_{s \neq s^*}-\text{MSE}\bigg\lvert}{\bigg\rvert\text{MSE}\big\rvert_{s \neq s_{\text{Best}}}-\text{MSE}\bigg\lvert},
\end{equation}
where
$\bar{N}(s^*)$ denotes the mean number of residues across the entire dataset. In this way, the importance $I(s^*)$ is given as a percentage of $s_\text{Best}$, the region that, when excluded, causes the MSE to deviate the most, \textit{i.e.}, it contributes the most to an accurate prediction of $K_D$. Since targets of different types (\textit{e.g.}, proteins or peptides) appeared to be associated to different binding configurations, we computed antibody region importance $I(s^*)$ for each target type $t$. Thus, the mean squared error in equation \ref{eq:importance_factor} was given by 
$\text{MSE} = \frac{1}{A_t}\sum_{i=0}^{A_t-1}( \log_{10}(K_{D}^{(i)})-f_{\theta^{*}}(\mathbf{X}^{(i)}))^2$, where
$A_t$ is the number of antibodies in the training set that bind to target type $t$, and $f_{\theta^{*}}$ denotes the binding affinity prediction by the final ANTIPASTI architecture with learned parameters $\theta^{*}$.

\begin{figure}
\centering
\includegraphics[clip, trim=0cm 0.8cm 9.5cm 0cm, width=\linewidth]{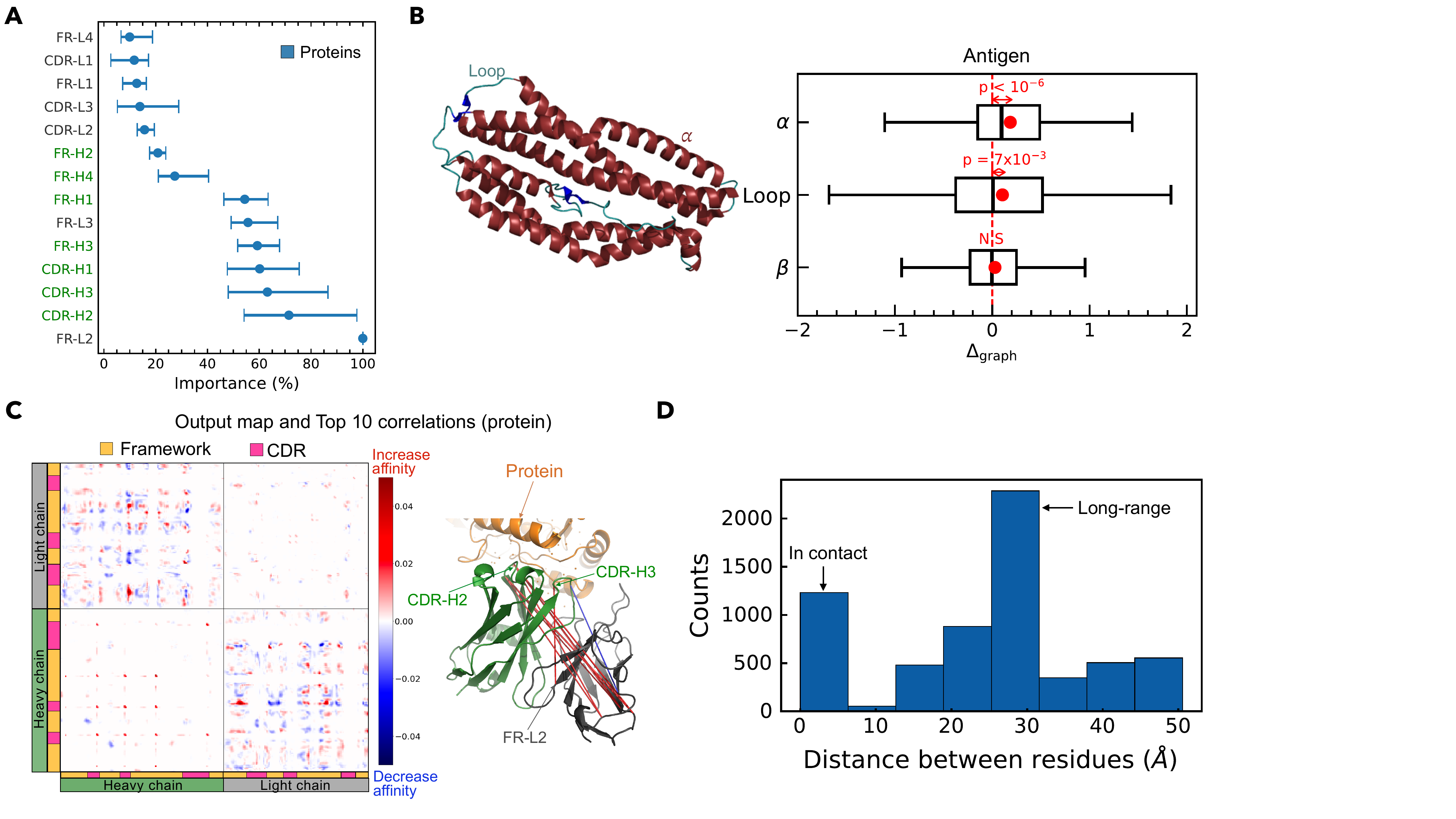}
\caption{\textbf{Biophysical interpretation in ML models.} (\textbf{A}) Ranking of ANTIPASTI importance factors for different antibody regions for protein targets. The importance factor (equation~\ref{eq:importance_factor}) is 
expressed as a percentage of that of the best region, and we show the average (dot) and extreme values (error bars) over 5 training/test splits. (\textbf{B}) Boxplot of INFUSSE's $\Delta_{\mathrm{graph}}$ stratified by secondary structure types ($\alpha$-helix, $\beta$-strand or loop) for the antigens in the test set. (\textbf{C}) ANTIPASTI's affinity-relevant correlations for an antibody with a protein target (Mntc with Mab 305-78-7 complex, PDB entry: \texttt{5hdq}). (\textbf{D}) Distribution of $\alpha$-carbon pairwise distances involved in the top $10$ ANTIPASTI's affinity-relevant correlations across all antibody structures. Panel B adapted from Ref.~\cite{Michalewicz2025}; panels A, C and D adapted from Ref.~\cite{Michalewicz2024}.}
\label{fig:5}
\end{figure}

We computed the region importance $I(s^*)$ for the FRs and CDRs across the heavy and light chains. For example, for antibodies binding to protein targets (Figure~\ref{fig:5}A), the three CDR loops of the heavy chain rank among the top four regions, consistent with their harboring most epitope-binding residues~\cite{Koenig2017,Akbar2022}. The CDR loops of the heavy chain are surpassed only by the second framework of the light chain (FR-L2). Previous experimental work has shown that mutations in FR-L2 can lead to improved thermostability, a pre-requisite for effective binding~\cite{Koenig2017}. Further, residues within FR-L2 are part of the heavy-light chain interface involved in highly stabilizing interactions with the CDR-H3, thus modulating its rigidity~\cite{Xu2015,Phillips2023}.

\paragraph{\normalfont{\textbf{Importance of structural information for B-factor prediction.}}} Within INFUSSE we perform a different type of ablation analysis that assesses the importance of structural information for B-factor prediction. Specifically, INFUSSE quantifies the effect of incorporating structural information through graph representations by comparing per-residue squared errors before and after adding the graph block $G_{block}$:
\begin{align}
\varepsilon^{(q)}_{S_{\mathrm{block}}, j} &:= \left(\hat{B}^{(q)}_{S_{\mathrm{block}},j} - B^{(q)}_j \right)^2 = 
\left([
    S_{\mathrm{block}} (\mathbf{x}^{(q)}_{\mathrm{s}}, 
 \mathbf{E}^{(q)}_{\mathrm{s}})]_j -B^{(q)}_j \right)^2,\\
    \varepsilon^{(q)}_{\text{INFUSSE}, j} &:= \left(\hat{B}^{(q)}_{j} - B^{(q)}_j \right)^2 = \left( [\text{INFUSSE}(\mathbf{x}^{(q)}_{\mathrm{s}}, \mathbf{r}^{(q)})]_j -B^{(q)}_j \right)^2,
\end{align}
where $j=1, \ldots, N_q$ indexes the residues of the PDB entries $q \in \mathcal{Q}$ in the test set. We then collect the differences in prediction errors into the set:
\begin{align}
\Delta_{\mathrm{graph}} =
\left \{  
    \Delta^{(q)}_{\text{graph},j}
    \right \} \quad \text{where} \quad 
   \Delta^{(q)}_{\text{graph}, j} &:= 
   \varepsilon^{(q)}_{S_{\mathrm{block}}, j}-\varepsilon^{(q)}_{\text{INFUSSE}, j} \; .
   \label{eq:Delta_fullset}
\end{align}
The distribution of $\Delta_{\mathrm{graph}}$ across residues enables us to assess whether specific antibody regions or structural motifs are associated to statistically significant performance gains when the structural graph information is added. For instance, we analyzed different secondary structure motifs ($\alpha$-helices, $\beta$-strands and loops) across antigens, and found that graph information brings an improvement for $\alpha$-helices and loops, but no significant change, compared to sequence-based prediction alone, for $\beta$-strands (see Figure~\ref{fig:5}B). We found a significantly higher $\Delta_{\mathrm{graph}}$ mean value for $\alpha$-helices compared to loops (difference of means $\Delta\mu=0.08$ with p-value $0.05$), while the interquartile range (IQR) for loops is wider than that of helices ($\Delta\mathrm{IQR}=0.25$ with p-value $10^{-6}$), suggesting more heterogeneity of graph-induced effects on performance across positions in loops. This aligns with our intuition: although graph information is beneficial to predictions in \emph{unstructured} regions, their intrinsic flexibility and conformational variability make the mapping between sequence, structure and B-factors less consistent in the training dataset, leading to more heterogeneous performance. On the other hand, the narrow, near-zero distribution of $\Delta_{\mathrm{graph}}$ for $\beta$-strands indicates that the sequence block is already able to capture most of their geometry and conformational flexibility, which are encoded to a large extent in their amino acid composition. $\beta$-strands are indeed particularly regular structural motifs whose stability is enabled by specific patterns of hydrophobic residues and hydrogen bonds~\cite{Wouters1995}.


\subsection{Model-Dependent Interpretability: Using the Learned Representations}
Model-dependent interpretability examines and probes the internal mechanisms of a model, such as parameter values or latent representations, resulting in explanations that are tied to the model architecture itself rather than being model-agnostic. Visualizing such internal parameters and representations, \textit{e.g.}, by projecting onto lower dimensions (see Figure~\ref{fig:2}D), provides complementary insight into the information that is encoded and leveraged for prediction at different layers~\cite{Lipton2016}. 


Beyond visualization, gradient-based perturbation methods use the value of gradients upon small perturbations of the inputs and their propagation to probe the predictive contribution of single neural network components (like hidden unit activations) or input features~\cite{Simonyan2014,Alipanahi2015}. DeepLIFT addresses key design limitations of seminal gradient-based perturbation methods by propagating difference-from-reference contributions to quantify input feature importance~\cite{Shrikumar2017}.

Within specific architectures, attention weights are another prototypical example, as the attention mechanism is designed to highlight which positions a model focuses on during prediction, and in protein language models these attention patterns have been shown to correlate with functional correlations and structural contacts~\cite{Vig2021}, including epitope-paratope ones~\cite{Leem2022}. Attention scores can also be recast through the lens of pairwise statistical mechanics models of proteins (Potts models), thereby linking learned dependencies to physical interactions between residues~\cite{Rende2024}. Finally, Transfer Learning applied learning network weights within the architecture of Restricted Boltzmann Machines can flag up distinctive amino acid patterns underlying molecular binding specificity in T-cell mediated immune responses~\cite{Bravi2023}. 

Within our methods, model-dependent interpretability has been employed in ANTIPASTI. Since the output layer of ANTIPASTI is designed such that the predicted binding affinity is a weighted sum of $M$ representational elements (equation \ref{eq:ba}), these can be arranged into an interpretable, 2D map $\mathbf{F}$ given by: 
%
\begin{equation}F_{jk}:= -\sum_{\ell=0}^{n_\text{f}-1}w_{i}z_{i}=F_{kj}, \quad \text{with} \quad i=j+\sqrt{\frac{M}{n_{\text{f}}}}k+\ell\frac{M}{n_{\text{f}}},
\end{equation} 
where $j,k=0,\dots,\sqrt{M/n_{\text{f}}}-1$ and $n_\mathrm{f}$ is the number of filters in the convolutional layer.
%
Through this reshaping, each element $F_{jk}$ can be interpreted as a representation of the NM-mediated correlation between amino acids $j$ and $k$ learned for the binding affinity prediction task.

Given that equation \ref{eq:ba} can be rewritten as 
$\log_{10}(\hat{K}_D) = -\sum_{j,k}F_{jk}$, 
the positive (respectively, negative) elements of $\mathbf{F}$ can be interpreted as the residue-residue correlations responsible for its high (respectively, low) binding affinity. We call these $F_{jk}$ pairs ``affinity-relevant correlations''. For instance, Figure~\ref{fig:5}C shows, among the ANTIPASTI's affinity-relevant correlations that are highest in magnitude, correlations linking two \textit{distant} regions: CDR-H3 and FR-L2. To further validate this insight that affinity-relevant correlations can span large distances in the structure, we considered the $10$ correlations highest in magnitude for each structure in the dataset. We found that most of the important correlations are long-range: $80\%$ of the top correlations are between residues more than $10\angstrom$ apart. Yet the distribution also has a peak at short distances, indicating that the model detects important short-range, contact-related physical interactions as well (Figure~\ref{fig:5}D). An open question amenable to future investigations is what biophysical processes underpin the long-range affinity-relevant correlations. In our preliminary inspection of a case where a 10-fold increase in binding affinity was observed upon a mutation on the CDR-H3~\cite{Schiele2015}, we found that the largest changes in the affinity-relevant correlations concerning the mutated residue connected it to two $\beta$-sheets that underwent substantial conformational shifts, as the mutation improved the steric fit of the antigen within the antibody's binding pocket (see Ref.~\cite{Michalewicz2024}). 

In both our models, the performance comparison to simpler approaches revealed that putative molecular determinants of the predicted properties entail structural and residue-residue interaction patterns rather than isolated amino acids. INFUSSE allows for the diffusion of information over local neighborhoods on geometry-derived graphs, capturing short- to mid-range motifs (\textit{e.g.}, loops, helices, or interfacial patches). ANTIPASTI emphasizes global dynamical couplings, as its affinity-relevant correlations are frequently inter-region and long-range. This is consistent with the well-acknowledged view that cooperative interactions and collective motions underlie many aspects of protein function~\cite{Eyal2011}, more than any single residue in isolation.

\section{Conclusion}\label{sec:conclusion}

In this chapter, we have examined how sequence and structure protein representations,
ML architectures and interpretability strategies can be tailored to antibody modeling and its specific challenges posed, for instance, by the functional constraints of antigen-binding specificity, by the scarcity of functionally annotated data, and by antibodies' sequence variability and conformational flexibility. We have argued that structure-derived representations, such as graphs, are particularly valuable in this context, as they can capture both local interactions and chemical neighborhoods, as well as long-range cooperative effects.

We have discussed two ML methods for antibody property prediction, ANTIPASTI and INFUSSE, both of which are underpinned by graph representations of protein structure, and which also illustrate the broader principle that ML predictive performance can be combined with interpretability for biological insight. 


ANTIPASTI provides an example of how binding affinity predictions (a \textit{global} property) can be linked to, and elucidated by, the direct inspection of learned representations as well as interpretable region-level importance factors. ANTIPASTI reveals that long-range cooperative correlations between regions, including contributions from the second framework of the light chain (FR-L2), can often be crucial to binding affinity. 

Our application of INFUSSE, on the other hand, focuses on B-factor predictions (a \textit{local}, residue-level property) and its modular architecture (one block trained on sequences, one block incorporating the graph representation) allows one to systematically assess the effect of incorporating structural information for improved accuracy of residue-level predictions. 
Systematic benchmarking against sequence-only and structure-only ablations confirms that coupling context-rich sequence embeddings with graphs improves accuracy, and pinpoints where graph information confers the greatest enhancement in prediction, notably within CDRs and at paratopes and epitopes. Due to its model-agnostic nature, the importance factor strategy used for ANTIPASTI could be extended to the local predictions by INFUSSE. All such quantitative analyses of region or residue importance to determining a given property could inform targeted antibody mutagenesis and ultimately accelerate antibody discovery.

Future work will be needed to embed such predictions into broader pipelines for computational antibody design by combining them with predictors of additional properties such as solubility, stability, and humanness~\cite{Rosace2023,Ramon2024,Ali2024}. This integration would allow antibody candidates to be assessed more holistically \textit{in silico}, with different predictors capturing orthogonal aspects of their molecular behavior. It remains however an open challenge to understand how to navigate the inevitable trade-offs between these properties in a systematic rather than an \textit{ad hoc} manner.

For design and more general purposes, a key step will be to create interpretability frameworks that can merge outputs across multiple models, connecting global and local predictions in a coherent way. This step would enable the development of ML models that not only score antibody candidates but also provide insight into the biophysical properties underlying their expected behavior, thereby enhancing model reliability and generalizability.
Another avenue for future work is the quantification of predictive uncertainty, in such a way that practitioners can gauge the extent to which they can rely on computational predictions ahead of experimental validation.

\end{document}